\documentclass[preprint]{aastex}

\usepackage{amssymb, amsmath, apjfonts, natbib, color}

\def\degree{{\circ}}
\newdimen\digitwidth
\setbox1=\hbox{0}
\digitwidth=\wd1
\catcode`"=\active
\def"{\kern\digitwidth}

\slugcomment{Astrophysical Journal Letters accepted}

\begin{document}

\title{ARE HIGH VELOCITY PEAKS IN THE MILKY WAY BULGE DUE TO THE BAR?}

\author{Zhao-Yu Li\altaffilmark{1}, Juntai Shen\altaffilmark{1, 6}, R. Michael Rich\altaffilmark{2}, Andrea Kunder\altaffilmark{3}, AND Shude Mao\altaffilmark{4, 5}}

\altaffiltext{1}{Key Laboratory for Research in Galaxies and Cosmology, Shanghai Astronomical Observatory, Chinese Academy of Sciences, 80 Nandan Road, Shanghai 200030, China}
\altaffiltext{2}{Department of Physics and Astronomy, University of California at Los Angeles, Los Angeles, CA, USA}
\altaffiltext{3}{Leibniz-Institute f\"{u}r Astrophysik Potsdam (AIP), An der Sternwarte 16, D-14482 Potsdam, Germany}
\altaffiltext{4}{Galaxy and Cosmology Division, National Astronomical Observatories of China, Chinese Academy of Sciences, A20 Datun Road, Room A505, Chaoyang District, Beijing 100012, China}
\altaffiltext{5}{Jodrell Bank Centre for Astrophysics, University of Manchester, M13 9PL, UK}
\altaffiltext{6}{Correspondence should be addressed to Juntai Shen: jshen@shao.ac.cn}

\begin{abstract}

Recently the commissioning APOGEE observations of the Galactic bulge reported that a significant fraction of stars ($\sim10\%$) are in a cold ($\sigma_{\rm V} \approx 30$ km/s) high velocity peak (Galactocentric radial velocity $\approx 200$ km/s). These stars are speculated to reflect the stellar orbits in the Galactic bar. In this study, we use two $N$-body models of the Milky Way-like disk galaxy with different bar strengths to critically examine this possibility. The general trends of the Galactocentric radial velocity distribution in observations and simulations are similar, but neither our models nor the BRAVA data reveal a statistically significant cold high velocity peak. A Monte Carlo test further suggests that it is possible for a spurious high velocity peak to appear if there are only a limited number of stars observed. Thus, the reported cold high velocity peak, even if it is real, is unlikely due to stars on the bar-supporting orbits. Our models do predict an excess of stars with high radial velocity, but not in a distinct peak. In the distance--velocity diagram, the high velocity particles in different fields exist at a similar distance $\sim8.5 \pm 1$ kpc away from the Sun. This result may be explained with geometric intersections between the line-of-sight and the particle orbits; high velocity stars naturally exist approximately at the tangent point, without constituting a distinct peak. We further demonstrate that even without the presence of a bar structure, particle motions in an axisymmetric disk can also exhibit an excess of high velocity stars.

\end{abstract}


\section{INTRODUCTION}

Near infrared (NIR) photometry revealed a boxy/parallelogram-shaped distortion in the Milky Way (MW) bulge region \citep{maih78, weil94, dwek95}, which has been suggested as a perspective effect: the near end of the bar is closer to the Sun than the far side \citep{blispe91}. As a normal disk galaxy, it is not surprising to find a bar structure in the Milky Way. Statistical studies reveal a high fraction of bars ($\sim60\%$) in disk galaxies \citep{marjog07}. As an effective mechanism to redistribute energy and angular momentum, the bar plays an important role in the secular evolution of disk galaxies \citep{korken04}.

$N$-body simulations have demonstrated that a dynamically cold disk can easily form a bar structure from internal instability, which subsequently thickens in the vertical direction due to the buckling perturbation to produce the observed boxy/peanut-shaped bulge in the edge-on view \citep{comsan81, raha91, mart06}. Both simulations and observations have confirmed that the side-on view of the boxy/peanut-shaped bulge usually demonstrates an X-shaped structure \citep{pfefri91, atha05, bure06}. The recently discovered X-shape in the Galactic bulge has also been shown to be related to the buckling process of the bar in the MW \citep{mcwzoc10, nata10, sait11, lishe12, ness12, cao13, wegger13}. The chemical analysis of the bulge region suggests that only the relatively metal-rich stars reveal the bar-related structure and kinematics \citep{babu10, hill11, ness12, utte12}. Although the detailed properties of the Galactic bulge such as the bar length, the tilted angle and the bulge-to-disk ratio are still unsettled (e.g., Sevenster et al. 1999; Bissantz \& Gerhard 2002; Bissantz et al. 2003, 2004; Babusiaux \& Gilmore 2005; Benjamin et al. 2005; Rattenbury et al. 2007; Martinez-Valpuesta \& Gerhard 2011; Gerhard \& Martinez-Valpuesta 2012), the existence of the bar and the peanut bulge in the MW is well recognized.

To better understand the MW bulge/bar kinematics and chemical evolution, detailed stellar spectroscopy is essential to obtain accurate radial velocity and element abundances. Recently, there has been a lot of progress in this direction. The kinematic analysis from the Bulge RAdial Velocity Assay (BRAVA) data displays a strong cylindrical rotation in the Milky Way bulge region, which is inconsistent with a spheroidal component  (Rich et al. 2007; Howard et al. 2008; Howard et al. 2009; Kunder et al. 2012). Shen et al. (2010) constructed a simple but realistic $N$-body model with a bar naturally developed from a dynamically cold disk. Once formed, this bar quickly buckles in the vertical direction due to the buckling/firehose instability (Raha et al. 1991). The major axis of the bar is $20^\degree$ away from the Sun--Galactic center (GC) line. From the solar perspective, the thickened part of the bar appears as the boxy bulge of our Galaxy, with the near side more extended than the far side. More quantitatively, as shown in Shen et al. (2010) and Kunder et al. (2012), this model well matches detailed stellar kinematics of BRAVA in all the observed fields.

Using the Sloan 2.5m telescope, the Apache Point Observatory Galactic Evolution Experiment (APOGEE) has been observing stars in infrared with an emphasis on red giants in the Galaxy (Eisenstein et al. 2011). The commissioning phase of APOGEE observed 18 bulge fields, which produced $\sim4700$ K/M giant stars in the Milky Way bulge. Based on the commissioning data towards the bulge region, Nidever et al. (2012) reported the discovery of a cold ($\sigma_{\rm V} \approx 30$ km/s) high radial velocity peak ($V_{\rm GSR} \approx 200$ km/s) in the Galactic bulge. They suggested that these high velocity stars can be best explained by stars on orbits of the Galactic bar potential. Although they verified the sample completeness and excluded other mechanisms like the tidal stream contamination, there are still unsettled questions about their conclusion. For example, the fields with the prominent high velocity peak are not symmetric about the Galactic plane, which is unexpected in a stable MW bar model. The peak itself is not statistically significant in some fields; at $(+4.3, -4.3)$, $(+5.7, -2)$ and $(+14, \pm 2)$, the high velocity peaks only contain $\sim$5 more stars than the adjacent trough at $\sim$150 km/s. Therefore, given the success of the model in Shen et al. (2010), we search for the existence of this high velocity peak in our simulation. In this process, we can also test its possible connection with the bar. An additional model with a strongly buckled bar is used to further probe the high velocity feature. Our models are briefly described in Section 2. In Section 3, the results of the comparison between the simulations and the observations are shown. We summarize our main results in Section 4.

\section{MODELS}

Model 1 comes from Shen et al. (2010), where one million particles evolve in a rigid dark matter potential. The initial exponential disk is dynamically cold (Toomre's $Q \approx 1.2$). As the simulation evolves, a bar quickly forms and buckles in the vertical direction. The structures of this simulated disk galaxy become steady after $\sim2.4$ Gyr. The snapshot we adopted here is at 4.8 Gyr, the same as Shen et al. (2010) and Li \& Shen (2012). The distance of the Sun to the GC ($R_0$) is about 8.5 kpc, and the Sun--GC line is $20^\degree$ away from the major axis of the bar. As already shown in Shen et al. (2010) and Li \& Shen (2012), this model successfully matches the observed kinematics and the X-shaped structure in the Galactic bulge. Additional details are given in Shen et al. (2010). For the purpose of comparison, we carried out another simulation (Model 2). In this case, disk particles evolve in a live dark matter halo, which enhances the growth of the bar to induce strong buckling instability. Initially, there are two million particles in the disk with an exponential density profile. The density distribution of the dark matter halo is described by an adiabatically compressed King profile ($\Phi(0) / \sigma^2 = 3$ and $r_{\rm t} = 10R_{\rm d}$, see Sellwood \& McGaugh\ 2005 for details of adiabatic compression). The halo consists of 2.5 million particles, and its total mass $M_{\rm halo} = 8M_{\rm d}$. The bar grows and quickly buckles in the vertical direction more strongly than Model 1. With the same solar position relative to the bar, this model is also scaled to match the BRAVA observations. More details about this model will be given in another work (Li et al. 2014, in preparation). The following analyses are unaffected if a smaller $R_0$ is adopted in the two models.

\section{COMPARISON WITH SIMULATIONS}

\subsection{Radial Velocity Distribution}

Figure~1 shows the comparison between the normalized $V_{\rm GSR}$ distributions from APOGEE commissioning data, Model 1 and Model 2 in the same fields. When producing these histograms from simulations, we only use particles with distances between 3 kpc and 13 kpc from the Sun to avoid the possible contamination from foreground and background particles. In Figure~1, the overall distributions agree quite well with each other in all the fields. Despite an excess of particles around 200 km/s, a separate high velocity peak is not seen in the models. For the BRAVA data in the similar fields, there is also no hint for such a statistically significant peak (see e.g., Figures~16, 17 and 18 of Howard et al. 2009, as well as Figures~13 and 15 of Kunder et al. 2002); the likely high velocity peak in some fields only contains about 5 stars. Similarly, in several APOGEE commissioning fields such as $(+4.3, -4.3)$, $(+5.7, -2)$ and $(+14, \pm2)$, there are only $\sim4-6$ more stars in the ``identified'' high velocity peak than the adjacent trough at slightly lower velocity.

In Nidever et al.\ (2012), the APOGEE commissioning radial velocity distributions were also compared with several model predictions, including Besan\c{c}on Galaxy Model (Robin et al. 2003, 2012), Kazantzidis $N$-body model (Kazantzidis et al. 2008) and the bar model used in Martinez-Valpuesta \& Gerhard (2011; hereafter MVG). In most bulge fields, these models do not show a prominent high velocity peak either (see their Figure~4). Although the authors claim good agreements with the Kazantzidis and MVG models, the velocity distributions of the two models still lack a clear trough around 150 km/s. In fact, our simulations predict similar velocity distributions to the Kazantzidis and MVG models.

From the solar perspective, the positive longitudinal regions from $4^\degree$ to $14^\degree$ correspond to the leading (near) side of the bar with most particles moving away from us ($V_{\rm GSR} > 0$). On the other side, negative longitudinal regions from $-10^\degree$ to $-4^\degree$ mainly contain the trailing (far) side of the bar with negative $V_{\rm GSR}$. If the cold high velocity peak claimed by Nidever et al. (2012) indeed exists, one should also expect a cold high velocity peak moving in the opposite direction in the negative longitudinal fields. Due to the lack of observations in the negative longitudinal area, APOGEE commissioning could not verify this argument. We carefully examine our models and BRAVA data in negative longitudinal regions  (Figure~13 of Kunder et al. 2012) and find no evidence for the cold high velocity peak in the opposite direction. Given the symmetric particle motions in the bar structure, it is hard to understand why the cold high velocity peak only appears in the leading side of the bar rather than the trailing side. Moreover, in a stable Milky Way model, the bar is expected to be symmetric with respect to the Galactic plane. This is apparently not the case for the fields in Nidever et al. (2012). In their Figure~2, the fields $(+4.3, -4.3)$ and $(+5.7, -2)$ are found with cold high velocity peaks, while the upper corresponding fields $(+4.3, +4.3)$ and $(+5.7, +2)$ do not show such a clear high velocity peak. Thus, based on our bar models, the observed cold peaks are difficult to explain with stars in the Galactic bar.

We find that small number statistics might have contributed to the appearance of such a high velocity peak. We perform a Monte Carlo (MC) test on our models. The same number of particles as the APOGEE commissioning observations are randomly selected from our models in each field of Figure~1. This process is repeated 10000 times. We find that a large fraction ($\sim 15 \%$) of these randomly selected subsamples can show a high velocity peak in the $V_{\rm GSR}$ distribution similar to observations\footnote{The position and height differences of the main peak, the trough and the high velocity peak between observations and simulations are required to be less than 30 km/s and 10 stars, respectively.}, even though the underlying true distribution is close to Gaussian. Examples of the subsamples in the field $(+14, +2)$ with a clear high velocity peak around $+200$ km/s are shown in Figure~2 with the underlying original distribution (shaded histogram) of Model 1. Note that the peak may be seen anywhere between $-250$ km/s and $+250$ km/s, but we select to show only those histograms with peaks near $+200$ km/s. This test has demonstrated that a cold high velocity peak can be produced, even though the underlying distribution is smooth. It is worth mentioning that observations with larger number statistics do not seem to find such a high velocity peak. Rangwala et al. (2009) measured radial velocities of 3360 red clump giants in three fields, i.e., the Baade's Window (+1.06, -3.9) and $(\pm5, -3.5)$. The radial velocity distributions of the three fields show no hint of a high velocity peak. Although small number statistics may be one possible reason, other factors, such as some unknown stellar stream(s) or co-moving group, might conceivably contribute to the peak as well. Of course, the relatively larger velocity dispersion close to the Galactic plane may also play a role here. If there are not enough stars in a field, the large velocity dispersion could give rise to spurious features especially in the tails of the distribution.

\subsection{Origin of High Radial Velocity Stars}

An essential parameter that can better link stars of different kinematics to structures in the MW is the distance. At different $l$'s, our line-of-sight actually passes through different structures inside the bar. The bar roughly extends to $|l| < 20^\degree$ in the longitudinal direction with the buckled region (the peanut structure) inside $|l| < 10^\degree$. For a sequence of bulge fields from $l = 2^\degree$ to $18^\degree$ on the disk plane ($b = 0^\degree$) in Model 1, we study the distribution of particles in the distance--$V_{\rm GSR}$ diagram (DVD) shown in Figure~3. The average $V_{\rm GSR}$ at different distances are shown with black lines. As $l$ increases, the particle distribution in DVD becomes less spatially concentrated. Also, the fraction of particles with negative radial velocities ($V_{\rm GSR} < 0$) drops from $\sim 40\%$ at $l = 2^\degree$ to $\sim 10\%$ at $l = 18^\degree$. 

In Figure~3, the fields with $l < 10^\degree$ (lower panels) are located inside the peanut structure, which is dense and dynamically hot. The spatial distribution of particles is concentrated at $\sim8.5$ kpc with a wide range of radial velocity (from $-150$ km/s to $+250$ km/s). In the upper panels, as the line-of-sight gradually moves outside the peanut structure, the particle distribution in DVD becomes spatially diffuse. The velocity distribution is also narrower suggesting less random motions, which reflects the kinematics of the outer unbuckled region of the bar.

Interestingly, in all DVDs from $l = 2^\degree$ to $18^\degree$, the maximum radial velocity speed can only be reached at distances around $R_0 \pm 1$ kpc from the Sun. The average velocity reaches maximum ($\sim 150$ km/s) at about 8 kpc, where a large fraction of high velocity particles reside. The average velocity is similar to the results in Babusiaux et al. (2014). At both smaller ($< 7.5$ kpc) and larger distances ($> 9.5$ kpc) in a given field, the maximum radial velocity speed drops. Assuming the bar is composed of co-aligned $x_1$ orbits of different sizes, the maximum $V_{\rm GSR}$ in each DVD can be found approximately at the distance where the line-of-sight is tangent to the orbits. Of course this is only approximately true since the speed along a bar-supporting $x_1$ orbit is not a constant value. The surface density profile of the simulated bars is roughly exponential; it is {\it highest close to the tangent point} along each line-of-sight, which increases the number of high velocity stars at that distance. As the line-of-sight shifts away from GC, there are naturally less particles with negative radial velocities due to stronger rotation of the disk and weaker random motions.

However, a bar is not necessarily required to produce these high velocity stars. To further illustrate this point, we perform a scrambling test; ($x, y, v_x, v_y$) of each particle in Model 1 are rotated by a random angle $\theta$ with respect to the GC. $\theta$ is uniformly distributed between 0 and $2\pi$. This results in a pure unbarred disk, which remains unbarred even if we evolve it with our $N$-body code. In Figure~4, the lower panel shows the original face-on image of Model 1, while the upper panel is the face-on image of the scrambled model. The Sun is marked with a black dot in the two panels. Different lines-of-sight are shown with dashed lines from $0^\degree$ to $20^\degree$ along the Galactic longitude. We carefully select the density contours to match the tangent point of each dashed line. As expected, after scrambling, the particles do not show any bar related structure or kinematics. DVDs of the scrambled model in different fields are shown in Figure~5. Compared to Figure~3, particles in the scrambled model show very similar distributions in DVDs, except that the distance of maximum radial velocity is slightly further away. This is also expected since the tangent point of a circular orbit is generally further from the Sun than the $x_1$ orbits for a given line-of-sight. This test demonstrates that particle motions in an axisymmetric disk can also produce similar results; a bar is not necessarily required.

\section{CONCLUSION}

With two self-consistent $N$-body models, we critically examine the cold high velocity peak towards the Galactic bulge, which was first reported in Nidever et al. (2012) with the APOGEE commissioning observation and suspected to reflect the stellar orbits in the Galactic bar. We carefully compare the radial velocity distribution of particles in our models with APOGEE commissioning and BRAVA observations towards the Galactic bulge. Despite an excess of stars at $\sim200$ km/s, i.e., an extended high velocity shoulder in the models, we do not find a prominent peak with small velocity dispersion. In addition, assuming that the bar structure is symmetric, we should expect a corresponding cold high velocity peak in the opposite direction at the  negative longitude. The radial velocity distributions of both our models and BRAVA do not show such a feature. Therefore, the reported cold high velocity peak, even if it is real, is unlikely due to stars on the bar-supporting orbits. According to our Monte Carlo test, it is possible for a cold high velocity peak to appear due to small number statistics. Observations with much larger number statistics thus far have not uncovered such a peak in the radial velocity distribution (e.g., Rangwala et al. 2009).

Our conclusion is in agreement with a parallel observational work using the APOGEE post-commissioning data (Kunder et al. 2014). They find that the two Nidever et al. (2012) high velocity fields re-observed after commissioning do not show a distinct high velocity peak. The most recent GIRAFFE Inner Bulge Survey (GIBS) obtained radial velocities of 6392 red clump giants in 24 bulge fields (Zoccali et al. 2014). In each field, they concluded that ``the radial velocity distribution shows no significant peaks (nor individual outliers) outside the main distribution, in contrast with the findings by Nidever et al. (2012)''.

We further investigate the bulge kinematics in our simulations with the DVD diagram. For the bulge fields within $|l| < 10^\degree$, the particle distribution in DVD is spatially concentrated with large fraction of negative radial velocities. In the outer part ($|l| \geq 10^\degree$), the distribution becomes diffuse with less contributions from random motions. One common feature in all DVDs is that the maximum radial velocity speed exists only at $\sim8.5\pm1$ kpc. As we show here, this can be explained with the location of these high velocity particles being the tangent point between the line-of-sight and the regular $x_1$ orbit in the bar. However, a bar is not necessarily required to produce these high velocity stars. Our scrambling test clearly demonstrates that particle motions in an axisymmetric unbarred disk can also show similar results. Thus the existence of these high velocity particles are naturally expected, which do not manifest themselves as a separate peak.

The research presented here is partially supported by the 973 Program of China under grants No. 2014CB845701, by the National Natural Science Foundation of China under grants No. 11073037, 11333003, 11322326, and by the Strategic Priority Research Program ``The Emergence of Cosmological Structures'' (No. XDB09000000) of Chinese Academy of Sciences. Hospitality at APCTP during the 7th Korean Astrophysics Workshop is kindly acknowledged.



\epsscale{1}
\figurenum{1}
\begin{figure}
\plotone{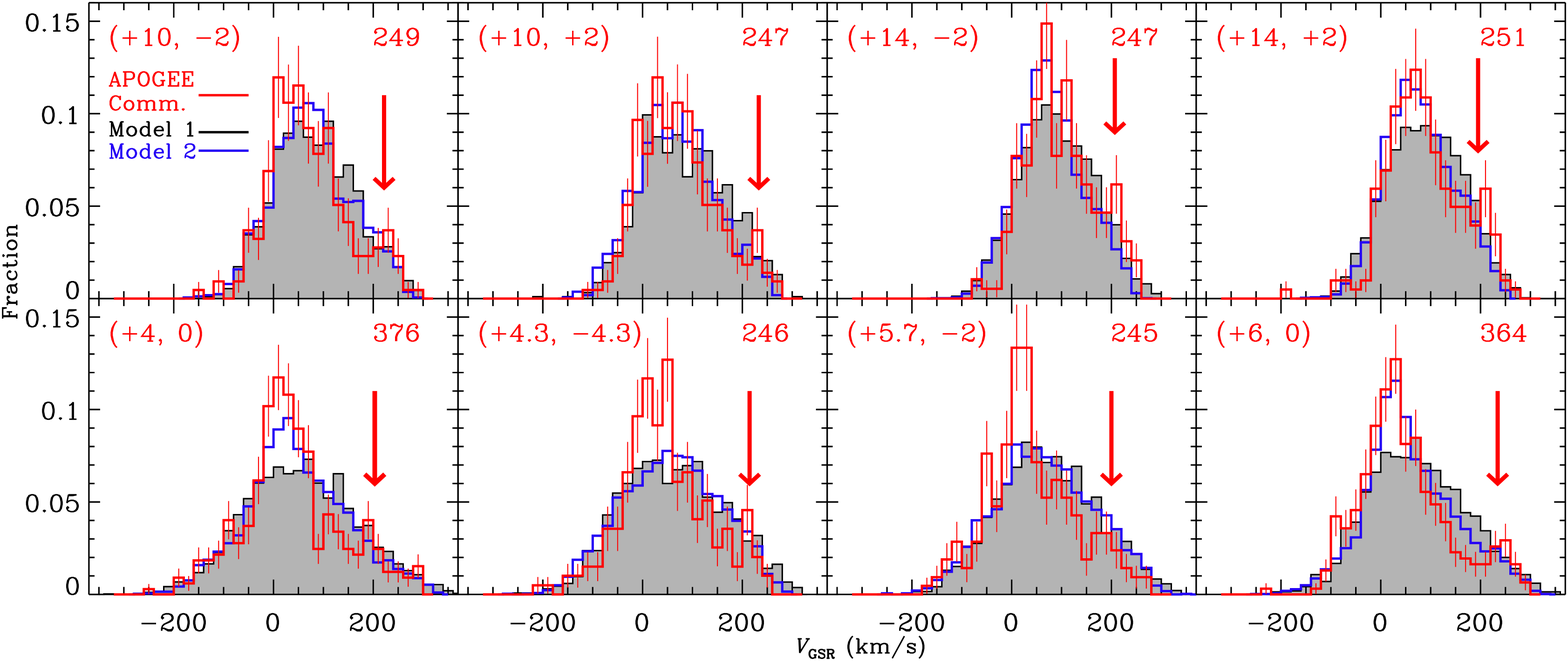}
\caption{Radial velocity ($V_{\rm GSR}$) distributions of Model 1 (grey shaded histogram) and Model 2 (blue histogram) in the same fields as Nidever et al. (2012) with the APOGEE commissioning observations overlaid (red histograms). The field name $(l, b)$ and number of stars are also given in each panel in red. The red arrows mark the positions of the high velocity peaks identified in Nidever et al. (2012).}
\end{figure}

\figurenum{2}
\begin{figure}
\plotone{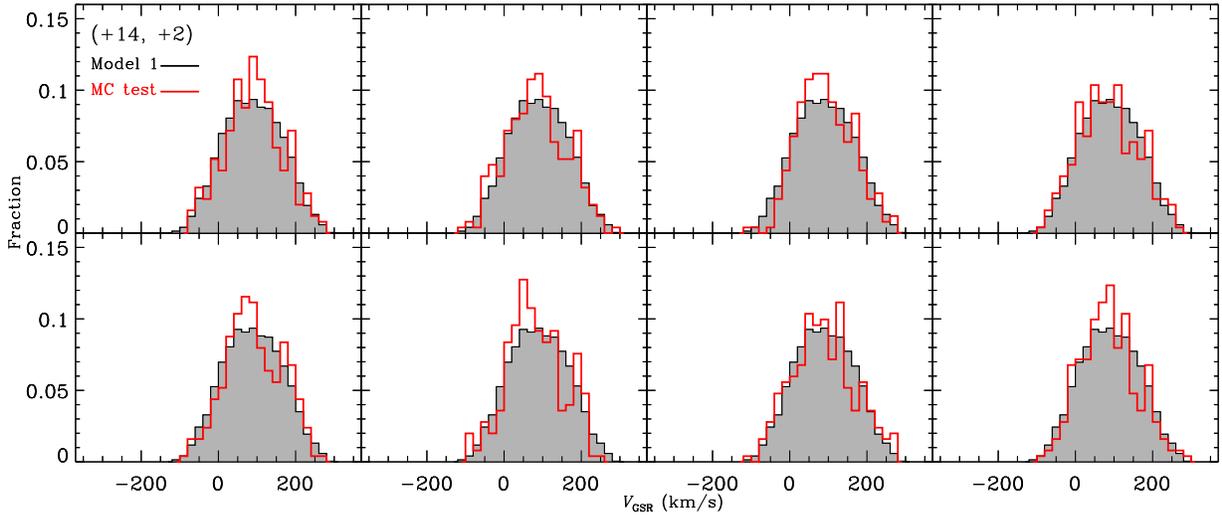}
\caption{Examples of the Monte Carlo subsamples with a high velocity peak in the $V_{\rm GSR}$ distributions in (+14, +2) for Model 1 (red histograms with 251 particles). The grey histogram shows the original $V_{\rm GSR}$ distribution of Model 1 in (+14, +2) with 2225 particles. We choose the same bin size (20 km/s) as in Nidever et al. (2012).}
\end{figure}

\figurenum{3}
\begin{figure}
\plotone{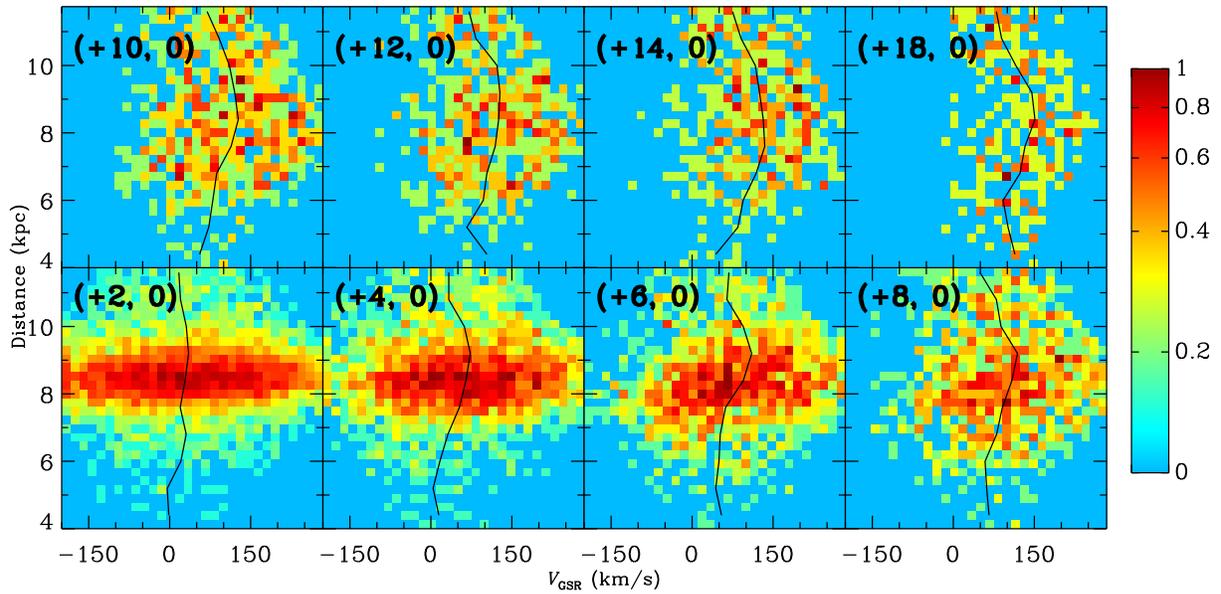}
\caption{The distance--$V_{\rm GSR}$ diagram of Model 1 in different fields towards the Galactic bulge. The ($l$, $b$) is given in each panel. Colors represent normalized number densities. The black solid line represents the average $V_{\rm GSR}$ at different distances.}
\end{figure}

\figurenum{4}
\epsscale{0.5}
\begin{figure}
\plotone{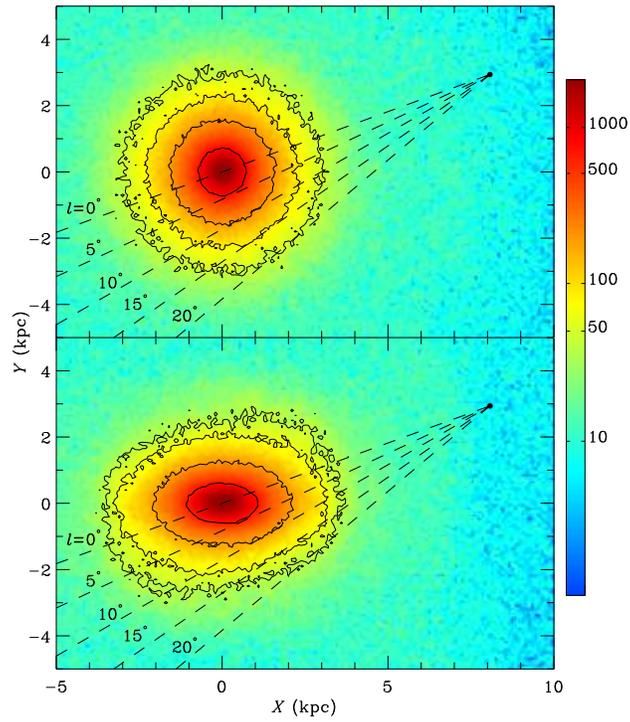}
\caption{Face-on views of Model 1 (lower panel) and the scrambled Model 1(upper panel). Particles rotate clockwise. The half length of the bar is $\sim4$ kpc. The Sun (black dot) is $\sim8.5$ kpc away from the GC, and the Sun--GC line is $20^\degree$ tilted from the major axis of the bar ($X$-axis). The dashed lines represent different Galactic longitudes from $0^\degree$ to $20^\degree$. Colors represent surface densities of the models in units of the number of particles per pixel.}
\end{figure}

\figurenum{5}
\epsscale{1}
\begin{figure}
\plotone{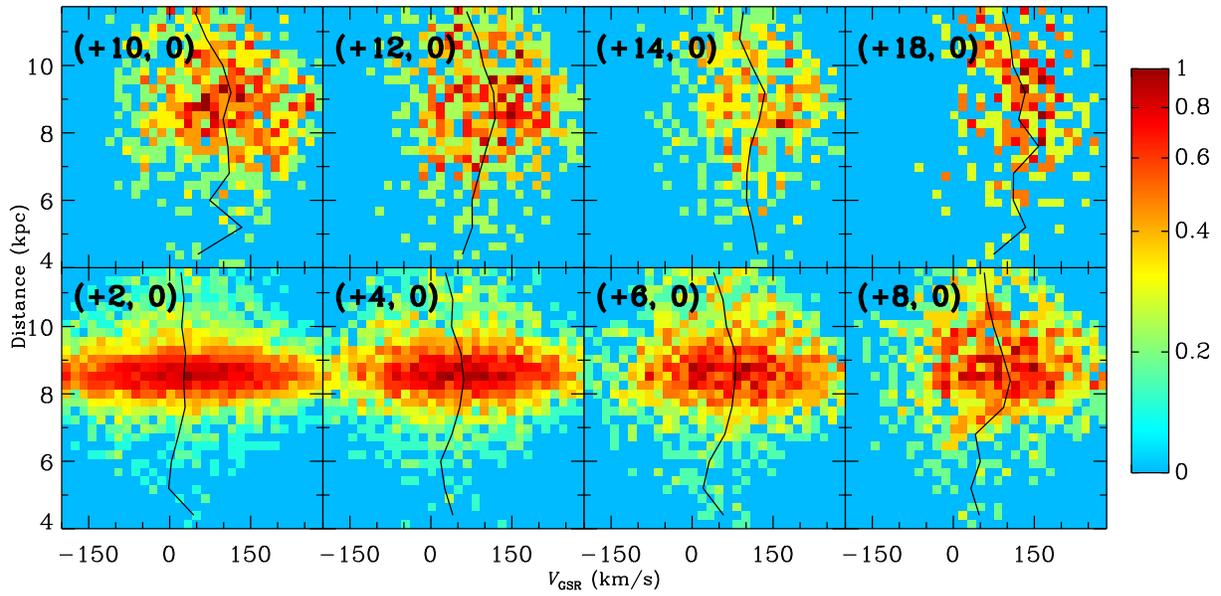}
\caption{The distance--$V_{\rm GSR}$ diagram of the scrambled Model 1 in the same field as Figure~3. The ($l$, $b$) is given in each panel. Colors represent normalized number densities. The black solid line represents the average $V_{\rm GSR}$ at different distances.}
\end{figure}

\end{document}